\documentclass[twocolumn,prl,aps,superscriptaddress,showpacs]{revtex4}
\usepackage{epsfig}

\begin{document}

\title{Chiral orbital-angular-momentum in the surface states of $Bi_2Se_3$}

\author{Seung Ryong Park} 
\affiliation{Institute of Physics and Applied Physics, Yonsei University, Seoul 120-749, Korea}
\affiliation{Department of physics, University of Colorado at Boulder, Boulder, Colorado 80309, USA}

\author{Jinhee Han}
\affiliation{Institute of Physics and Applied Physics, Yonsei University, Seoul 120-749, Korea}

\author{Chul Kim}
\affiliation{Institute of Physics and Applied Physics, Yonsei University, Seoul 120-749, Korea}

\author{Yoon Young Koh}
\affiliation{Institute of Physics and Applied Physics, Yonsei University, Seoul 120-749, Korea}

\author{Changyoung Kim}
\email[Electronic address:$~~$]{changyoung@yonsei.ac.kr}
\affiliation{Institute of Physics and Applied Physics, Yonsei University, Seoul 120-749, Korea}

\author{Hyungjun Lee}
\affiliation{Institute of Physics and Applied Physics, Yonsei University, Seoul 120-749, Korea}

\author{Hyoung Joon Choi}
\email[Electronic address:$~~$]{h.j.choi@yonsei.ac.kr}
\affiliation{Institute of Physics and Applied Physics, Yonsei University, Seoul 120-749, Korea}

\author{Jung Hoon Han}
\affiliation{Department of Physics and BK21 Physics Research Division, Sungkyunkwan University, Suwon 440-746, Korea}

\author{Kyung Dong Lee} 
\affiliation{Department of Physics, Inha University, Incheon 402-751, Korea}

\author{Nam Jung Hur} 
\affiliation{Department of Physics, Inha University, Incheon 402-751, Korea}

\author{Masashi Arita}
\affiliation{Hiroshima Synchrotron Radiation Center, Hiroshima University, Higashi-Hiroshima, Hiroshima 739-0046, Japan}

\author{Kenya Shimada}
\affiliation{Hiroshima Synchrotron Radiation Center, Hiroshima University, Higashi-Hiroshima, Hiroshima 739-0046, Japan}

\author{Hirofumi Namatame}
\affiliation{Hiroshima Synchrotron Radiation Center, Hiroshima University, Higashi-Hiroshima, Hiroshima 739-0046, Japan}

\author{Masaki Taniguchi}
\affiliation{Hiroshima Synchrotron Radiation Center, Hiroshima University, Higashi-Hiroshima, Hiroshima 739-0046, Japan}

\date{\today}

\begin{abstract}
Locking of the spin of a quasi-particle to its momentum in split bands of on the surfaces of metals and topological insulators (TIs) is understood in terms of Rashba effect where a free electron in the surface states feels an effective magnetic field. On the other hand, the orbital part of the angular momentum (OAM) is usually neglected. We performed angle resolved photoemission experiments with circularly polarized lights and first principles density functional calculation with spin-orbit coupling on a TI, $Bi_2Se_3$, to study the local OAM of the surface states. We show from the results that OAM in the surface states of $Bi_2Se_3$ is significant and locked to the electron momentum in opposite direction to the spin, forming chiral OAM states. Our finding opens a new possibility to have strong light-induced spin-polarized current in the surface states.
\end{abstract}
\pacs{73.20.At, 74.25.Jb, 71.15.Mb} \maketitle

Topological insulators (TIs) are a new class of materials and they distinguish themselves from ordinary insulators by their topological nature\cite{Fu, Moore}. Among the various properties of TIs, probably the most exciting one is that they always have metallic surface states because electrical conduction should occur in the metallic surfaces states\cite{Hsieh1, Zhang, Xia, Chen, Hsieh2, Park}. These surface states are topologically protected from perturbation by the time reversal symmetry. Moreover, electron spins of the surface states are locked into the momentum, making chiral spin structure\cite{Hsieh2}. This spin texture suppresses the back scattering\cite{Roushan, ZhangSTM,Alpichshev} and thus promotes possibility for TIs to be used for spin conserving media in spintronics\cite{Qu}. Locking of the electron spin to the momentum comes from a combination of strong spin-orbit interaction (SOI) and inversion symmetry breaking at the surface. This effect, known as Rashba effect, occurs on the surfaces of many materials and produces spin degeneracy lifted surface states\cite{Rashba}, such as Au(111) surface states\cite{Reinert}. 

While electron chiral spin texture is well studied, orbital angular momentum (OAM) is in many cases assumed to be quenched.  Indeed, in recent studies of TIs, surface states are found to have approximately $p_z$ character for which the OAM is quenched due to crystal field splitting\cite{Zhang}. Even though this may be a reasonable assumption for ordinary materials, there must be some tendency to restore a local OAM in the high-Z materials with strong SOI, and the local OAM must also have chiral structure by symmetry. Existence of chiral OAM in the surface states is important because it can provide electrostatic energy through electric dipole and surface electric field interaction\cite{Park2}. This is a crucial feature in the $quantitative$ description of Dirac cone-like surface state dispersion\cite{Park2}, while topological nature of bulk states determine $qualitative$ structure of the surface states only\cite{Fu, Zhang}. Chiral OAM is also important in optical properties of TI, since orbital part of the total wave function directly couples to electric field of light. Details in the importance of chiral OAM for optical properties will be discussed later.

In order to investigate OAM in the surface states of TI, we measured circular dichroism (CD) in angle resolved photoemission spectroscopy (ARPES) on $Bi_2Se_3$, one of the widest band gap topological insulators\cite{Zhang, Xia}, as CD in APRES has been proposed as a tool to measure OAM\cite{ParkJH}, and performed first principle caculation on the material. In this Letter, we show strong CD in ARPES from the surface states of $Bi_2Se_3$, indicating that the surface states possess substantial chiral OAM. First principles calculation also reveals chiral OAM with a sizable magnitude, proving the conclusion extracted from the experimental results. We discuss light induced strong spin (and also OAM) polarized current from the surface states which is directly deduced from existence of substantial chiral OAM.

\begin{figure}
\centering \epsfxsize=8cm \epsfbox{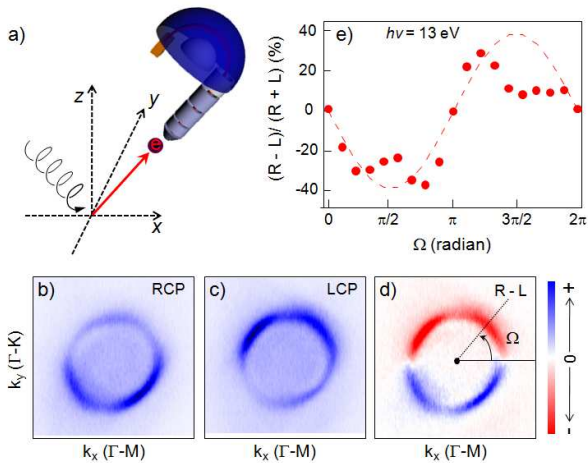} \caption{(a) Experimental geometry used in the study. $Bi_2Se_3$ surface is parallel to the $xy$-plane. Circularly polarized photons represented by the helical arrows travel in the $xz$-plane and are incident at an angle of 40 degrees to the sample surface. Fermi surface maps with 13 eV(b) RCP and(c) LCP lights as well as (d) their difference (RCP-LCP) are plotted. Azimuthal angle $\Omega$ is defined as the angle from the positive $k_x$-axis. (e) CD in ARPES at the Fermi level along the Fermi surface as a function of  $\Omega$. Difference between RCP and LCP is normalized by the sum of them. The dashed curve is the best sine function fits of the experimental data.}\label{fig1}
\end{figure}

Single crystals of $Bi_2Se_3$ were grown by a self-flux technique, following the previously reported recipe\cite{Hor}. ARPES measurements were performed at the beamline 9A of HiSOR equipped with VG-SCIENTA R4000 analyzer. Data were taken with left and right circularly polarized 10 eV and 13 eV photons. The total energy resolution was set to be 10 meV, and the angular resolution was $0.1^{\circ}$. Samples were cleaved in situ and the chamber pressure was about $5\times10^{-11}$Torr. The measurement temperature was kept at 15 K.
First-principles density functional calculations for the electronic structure of $Bi_2Se_3$ are based on $ab$ $initio$ norm-conserving pseudopotentials\cite{Troullier} and the Perdew-Burke-Ernzerhof-type generalized gradient approximation\cite{Perdew}, as implemented in the SIESTA package\cite{Portal}. We included the spin-orbit coupling in the SIESTA. A supercell is used to obtain the electronic structure of a slab of $Bi_2Se_3$ which consists of ten quintuple layers\cite{ZhangDFT}.

Depicted in Fig. 1(a) is the experimental geometry. $Bi_2Se_3$ surface is in $xy$-plane. Circularly polarized lights come in at 40 degrees to the $xy$-plane in the $xz$-plane. Photoemission intensity from the surface states is recorded as functions of energy and momentum for right- (RCP) and left-circularly polarized (LCP) lights. In Figs. 1(b) and (c), we plot ARPES intensity at the Fermi level as a function of the momentum for the data taken with $h\nu$ = 13 eV. The Fermi surface (FS) is where the ARPES intensity is high (blue color) and it has a hexagonal shape, consistent with published data\cite{Kuroda}. The skewness in the FS shape is an experimental artifact but does not affect our discussion. As for the intensity, we note that it is not uniform along the FS, which means that the photoemission matrix element has momentum dependence. Comparing the two data sets, we see a very clear difference between the two FSs taken with two different circular polarizations. We see higher intensities on the negative $k_y$ side for RCP but on the positive $k_y$ side for LCP lights. To see CD more clearly, we subtract LCP data from RCP data and plot the difference data in Fig. 1(d). The intensity profile of the FS is anti-symmetric about the horizontal axis ($k_y = 0$); $+k_y$ side is negative while $-k_y$ side is positive. For a quantitative analysis, we define CD by the difference in the ARPES intensities taken with two polarizations normalized by the sum of the two, that is, $(RCP-LCP)/(RCP+LCP)$. We plot in Fig. 1(e) CD at the Fermi level as a function of azimuthal angle $\Omega$. Also plotted in the figure are the best sine curve fit to the experimental data. One can see that CD is as large as about $30\%$ at $\Omega = \pi/2$ and $3\pi/2$. 

\begin{figure}
\centering \epsfxsize=6.5cm \epsfbox{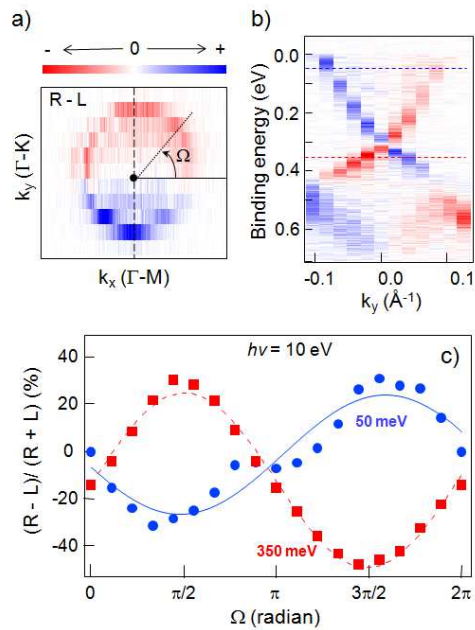} \caption{(a) RCP - LCP data taken with 10 eV photons. (b)$E$ vs. $k$ ARPES data along the dashed line in panel (a) ($k_x$= 0 cut).(c) CD in ARPES as a function of $\Omega$ for 50 (blue circle) and 350 (red square) meV data. The energy positions are marked by the dashed lines in panel (b). Difference between RCP and LCP is normalized by the sum of them and the lines are the best sine function fits of the experimental data.}\label{fig2}
\end{figure}

It is informative to see the binding energy dependent behaviour of the CD from the surface states. We found that low energy photons have to be used to separate out the surface state information by suppressing the bulk intensity\cite{Park}. We used 10 eV photons to take CD data. Similar to Fig. 1(d), we plot in Fig. 2(a) the difference data $(RCP-LCP)$ at the Fermi energy. Even though it has less number of momentum steps along the $k_y$ direction compared to the 13 eV data, the dichroic behaviour at the Fermi energy is roughly similar.  
The dichroic behaviour is the most prominent along the $k_y$-axis (dashed line in Fig. 2(a)). We plot the $E$ vs. $k$ data along the dashed line in Fig. 2(b). Dirac conelike bands of the surface states with the Dirac point at ~0.3 eV are seen as expected. On the other hand, the intensity behaviour of the two bands is opposite; the negative slope side has positive value (blue) while the positive side has negative value (red). Other than the Dirac conelike bands, states between 0.5 and 0.7 eV show CD but no sign of CD is observed for the states beyond 0.7 eV. 
We plot in Fig. 2(c) CD at 50 and 350 meV binding energies as a function of azimuthal angle $\Omega$ defined in Fig. 2(a). CDs for the binding energies above and below the Dirac point oscillate with a periodicity of $2\pi$ but are out of phase by $\pi$. Also plotted in the figure are the best sine curve fits to the experimental data. One can see that CD is as large as $30\%$ at $\Omega = \pi/2$ and $3\pi/2$.

What is causing the observed CD? In some cases, CD in ARPES can result from chiral experimental geometry effect\cite{Dubs, Westphal}. However, phase shift of $\pi$ between the data below and above the Dirac point as well as almost constant dichotic signal regardless of the distance from $\Gamma$ point shown in Fig. 2 cannot be explained by a simple chiral experimental geometry effect\cite{Dubs, Westphal}. In addition, CD in ARPES as large as $30\%$ is not expected in that case. 

\begin{figure}
\centering \epsfxsize=8cm \epsfbox{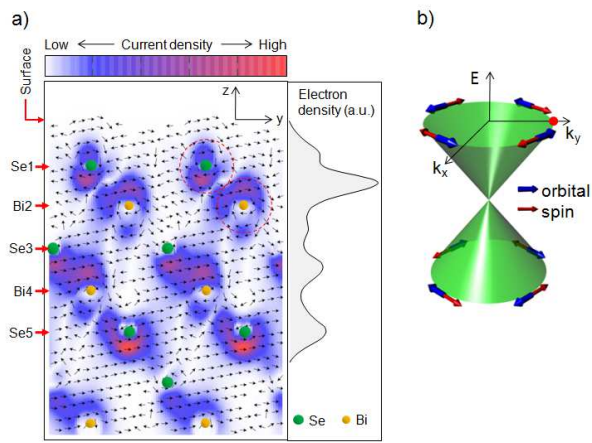} \caption{(a) Crystal structure of $Bi_2Se_3$, showing the first quintuple layer. Directions and magnitudes of the electron current density of a surface state are shown by arrows and in a colour scale, respectively. Shown on the right is the depth distribution of the surface state electron density. (b) Schematic band dispersion and OAM/spin configuration of surface states.}\label{fig3}
\end{figure}

We argue that the observed CD is due to existence of large chiral OAM in the surface states\cite{ParkJH}. The incident light at 40 degrees is a linear sum of light propagating in  the $xy$-plane and $yz$-plane. First, let us focus on the component in the $xy$-plane propagating along the $x$-axis. Such a light can deliver +1 OAM (RCP) or -1 (LCP) to the magnetic quantum number ($m_x$) of the electronic state due to the dipole selection rule. Let us assume that the initial state at $\Omega = 3\pi/2$ has $m_x = 1$. This state will be projected to $m_x=2$ and $m_x=0$ states by RCP and LCP, respectively. The difference between transition matrix elements for $m_x=2$ and $m_x=0$ photoelectron final states should result in CD. The CD can be shown to be roughly proportional to $\hat{m}\cdot\hat{k}_{ph}$, where $\hat{m}$ is the OAM direction and $\hat{k}_{ph}$ is the incoming photon direction (See Ref. \onlinecite{ParkJH} for details on the relationship between CD and OAM.).  The sinusoidal CD in Fig. 1(e) and 2(c) shows that the OAM direction has a chiral structure and is perpendicular to momentum as in the case of spins. In addition, one can find from Fig. 2(c) that the upper and lower Dirac cones have opposite OAM chirality.

We take further step to directly prove the existence of chiral OAM and performed first principle calculations within generalized gradient approximation on a $Bi_2Se_3$ slab. The overall band structure is consistent with the published result\cite{Zhang}. In Fig. 3(a), we plot the crystal structure of the first $Bi_2Se_3$ quintuple layer along with the real space electron density of the surface state as a function of the distance from the surface. Note that the electron density is not localized at the very surface atomic layer but is distributed over the first quintuple layer. 

\begin{table}
\centering \epsfxsize=8cm \epsfbox{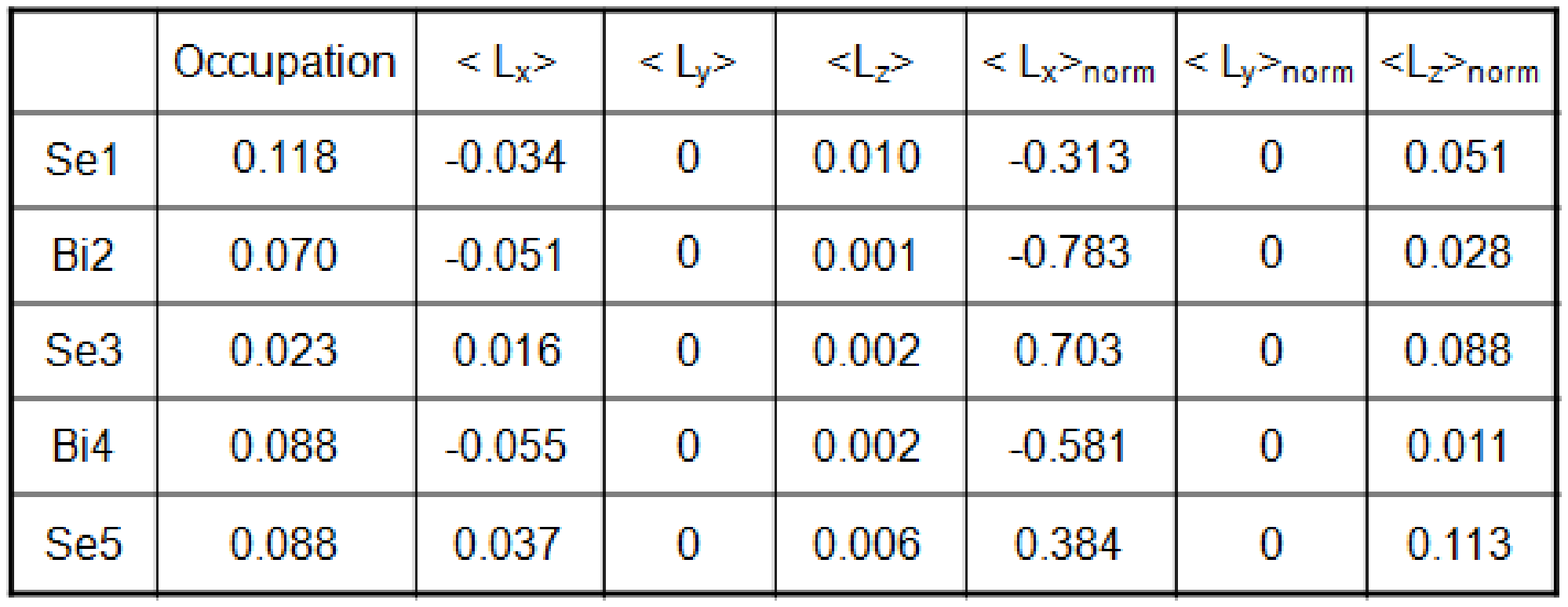} \caption{Listed are electron occupation, OAM and normalized OAM values for the surface state at $\Omega=\pi/2$. Values are calculated by projecting the surface state on $p$ orbitals of each atom. Angular momentum is given in units of $\hbar$.}\label{tab1}
\end{table}

We evaluate the local OAM of a surface state around each atom in the first quintuple layer by calculating the expectation values of the angular momentum operators using the projection of the surface state wave function on three $p$ orbitals of each atom. Table 1 lists calculated values at each atomic site. Column 1 is the electron occupation of the three $p$ orbitals obtained from the square of the projected wave function, which is the largest for Se1 and the smallest for Se3. Note that occupation numbers only sum up to 0.387 because they count the surface state wave function only in regions near the atoms in the first quintuple layer. Columns 2 to 4 list OAM expectation value calculated with the surface state at $E_F$ and $\Omega=\pi/2$ (marked by the red dot in Fig. 3(b)) in unit of $\hbar$. OAM is indeed non-zero, and is especially large for Bi atoms while values for Se atoms are small or even have the opposite sign. Since the major contribution comes from Bi atoms, we look at the values for Bi atoms. Note that orbital angular momenta for Bi atoms are pointing in the negative $x$-direction, which confirms our CD results. It is also noted that it is opposite to the spin direction. The expectation values may also be normalized by the electron occupation listed in column 1. Normalized values may be a good measure of the OAM strength, with the maximum value of $\hbar$ for a $p$ electron. They are listed in columns 5 to 7. Values for Bi atoms are large as expected, close to 80 $\%$ of the full value of $\hbar$.

Even though the expectation values listed in table 1 already confirm the existence of OAM, we can also visualize it. Plotted in Fig. 3(a) is the current density for the surface state at $E_F$ and $\Omega=\pi/2$ (represented by the red dot in Fig. 3(b)). The arrows represent the direction of the current while the density is shown in the colour scale. Overall, the current is flowing to the positive $y$-direction as expected. Looking at more details, one finds that the current rotating locally around atoms and it does in the direction that produces an OAM in the negative $x$-direction (except Se3 and Se5 atoms). All these results prove that OAM of the surface states in $Bi_2Se_3$ form chiral states.


\begin{figure}
\centering \epsfxsize=8.7cm \epsfbox{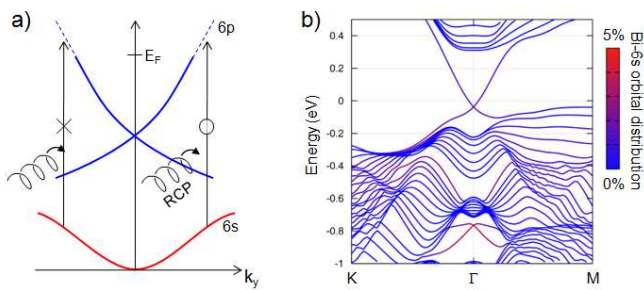} \caption{(a) With RCP lights, transition from 6$s$ states to surface states is allowed on the positive momentum side where $m_x=1$ while inhibited for the negative momentum side ($m_x=-1$), resulting in currents with spin polarization in the x-direction. (b) Distribution of the Bi2 6$s$ orbital in the band structure is shown in a color scale. The band structure is from a 10 quintuple layer slab calculation.}\label{fig4}
\end{figure}

An important implication of the large OAM in the surface states is that very strong spin (and also OAM) polarized current can be induced by light in surface states. This is expected to be, in principle, stronger than light induced spin-polarized current measured in binary semiconductor quantum wells\cite{Ganichev}, since the surface states in TI have single band crossing Fermi energy as well as very strong orbital polarization. From the dipole section rule, an electron with $s$-orbital character can be excited to the surface states ($p$-orbitals) by a photon. In addition, existence of OAM allows us to use circularly polarized photons to excite electrons only in certain part of the Brillouin zone. As the surface states are spin-polarized, selective excitation of electrons should result in spin current. Fig. 4(a) schematically shows the process. RCP photons, for example, excite electrons with positive momentum, resulting in current in the positive direction with spins-polarized along the $x$-direction. Then, the question is whether we have states with $s$-orbital character. Fig. 4(b) shows in a colour scale the contribution from 6$s$ orbital of Bi2 atom to the near $E_F$ states. Although Bi 6$s$ states mostly reside in the valence band around 10eV binding energy\cite{Mishra}, a small portion of them exists near $E_F$ as shown in Fig. 4(b). Especially, bands at about 0.6 and 0.8 eV binding energies near $\Gamma$ have about $3\%$ and $5\%$ of Bi 6$s$ orbital components, respectively. 

We calculated the transition rates between the two states shown in Fig. 4(b) for 0.7 eV LCP and RCP lights (equivalently, transition rates for $k$ and $-k$ states with the same polarization). The ratio between the two transition rates is found to be about 3.5, meaning that RCP lights will induce 3.5 times more current in the $k$ direction compared to $-k$ direction. Therefore, very strong light-induced spin and OAM polarized current can be realized in the surface states of TIs. It is recently reported that charge transport dominantly occurs in the surface states for high quality TI thin films\cite{Bansal}, and thus light induced spin and OAM polarized currents became more realistic. It could then be used for opto-spintronic applications\cite{Sarma}.

\acknowledgements
Experimental work is supported by the KICOS in No. K20602000008 and by Mid-career Researcher Program through NRF grant funded by the MEST (No. 2010-0018092). Computation part was supported by the NRF of Korea (Grant 2011-0018306) and computational resources have been provided by KISTI Supercomputing Center (Project No. KSC-2011-C2-04). ARPES measurements were performed with the approval of the Proposal Assessing Committee of HSRC (Proposal No.09-A-49).

\end{document}